# Automatic Classification of Bug Reports Based on Multiple Text Information and Reports' Intention

Fanqi Meng, Xuesong Wang[✉], Jingdong Wang[✉], and Peifang Wang

School of Computer, Northeast Electric Power University, Jilin City, Jilin, China
{mengfanqi,wangjingdong}@neepu.edu.cn, wxs970705@163.com

**Abstract.** With the rapid growth of software scale and complexity, a large number of bug reports are submitted to the bug tracking system. In order to speed up defect repair, these reports need to be accurately classified so that they can be sent to the appropriate developers. However, the existing classification methods only use the text information of the bug report, which leads to their low performance. To solve the above problems, this paper proposes a new automatic classification method of bug reports. The innovation is that when categorizing bug reports, in addition to using the text information of the report, the intention of the report (i.e. "suggestion" or "explanation") is also considered, thereby improving the performance of the classification. First, we collect bug reports from four ecosystems (Apache, Eclipse, Gentoo, Mozilla) and manually annotate them to construct an experimental data set. Then, we use Natural Language Processing technology to preprocess the data. On this basis, BERT and TF-IDF are used to extract the features of the intention and the multiple text information. Finally, the features are used to train the classifiers. The experimental result on five classifiers (including K-Nearest Neighbor, Naive Bayes, Logistic Regression, Support Vector Machine and Random Forest) show that our proposed method achieves better performance and its F-Measure achieves from 87.3% to 95.5% .

## 1 Introduction

Defect repair has an important impact on software quality assurance. It is the main activity in the later maintenance phase of software engineering. In recent years, with the vigorous development of the software engineering industry, the architecture complexity and code capacity of software systems have reached a new level that makes it difficult for developers to understand and manage [1]. This trend leads to a large number of bugs inevitably generated in the development process of software systems. To fix these bugs, developers must check the bug report [2]. The bug report describes the defects of the software system in the form of text, which contains multiple tags such as ID, Reporter, Summary, etc., as shown in Fig. 1. In the past, managers classified bug reports based on tags so that they could assign the reports to appropriate developers to accomplish bug fixes. However, this is a very time-consuming task as there are too many bug reports to check manually. Moreover, due to the different experience and knowledge background of the reporter, the report submitted in the Bug Tracking System may have incorrect tags [3]. These wrong tags will cause the bug report to not be correctly assigned to the appropriate developers, thereby increasing the difficulty of defect repair [4, 5]. In order to reduce this impact and accelerate the speed of defect repair, the software engineering industry needs accurate and automated classification methods for bug reports.

**Fig. 1.** Several examples of bug report from Bugzilla

In recent years, many researchers have explored the automatic classification of bug reports. Among them, Antoniol et al. [6] classified bug reports via text mining technology. It proved that automatically classify reports into bug and other types through training models is effective and feasible. Zhou et al. [7] proposed a hybrid method combining text mining and data mining techniques to determine whether a new bug report is a real bug. This method also considers the structured information of the report on the basis of purely mining text description [5] (Such as severity and priority). Lamkanfj et al. [8] adopted machine learning technology to classify bug reports into severity and non-severity. Similarly, Tian et al. [9] proposed a nearest neighbor solution based on information retrieval to predict the severity of the bug report. They focused on predicting the five severity levels of the report, namely: Blocker, Critical, Major, Minor and Trivial. In addition, some scholars are concerned about the quality of bug reports [10] and the imbalance of data sets [11, 12] and other issues.

Reporters submit reports with clear intentions. After reading the summary of a large number of open source software bug reports, we found that the intention of the summary text content can be classified into two types: explanation or suggestion. However, there is no intention label in the Bug Tracking System. A large number of existing studies fail to consider the intention of the report when classifying bug reports, which lead to lower performance of their methods. Considering that these intentions will affect the classification of reports, the method in this paper incorporates the intentions of the report. Among them, the explanation refers to the description of the defect, such as a problem or the cause of the problem in a certain location, and the suggestion refers to a solution to the defect, such as how to deal with a certain problem. Table 1 shows real examples of software bug reports in four different ecosystems and their intention.

**Table 1.** Bug reports with intention tags

| Ecosystem | Bug ID | Summary | Intention |
|---|---|---|---|
| Apache | 63099 | Regression in JMeter 5.0 due to fix of Bug 62478 | Explanation |
| Eclipse | 82281 | logical structures table should sort on name | Suggestion |
| Gentoo | 76636 | Kernel module dvb-ttpci does not find its firmware | Explanation |
| Mozilla | 277324 | Copy XML doesn't work on #document nodes | Explanation |

To sum up, the contributions of this paper are as follows:

(1) A new bug report classification method is proposed. Based on the text field classification, the intention of the report is additionally considered, and the report is classified as bug and no-bug.
(2) An automatic classification model was constructed based on the proposed method, and the classification performance of five classifiers (K-NN, NB, LR, SVM, RF) was observed. To measure the performance, the Accuracy, Precision, Recall, and F-measure are calculated.
(3) A dataset containing the intention and type of report that can be used by researchers to further explore the automatic classification of bug reports.

The rest of this paper is as follows: The Sect. 2 introduces the related work of bug report classification. The Sect. 3 introduces the proposed method. The Sect. 4 shows the experiment. The Sect. 5 discusses the experiment. Finally, the Sect. 6 summarizes the full text and puts forward insights on future work.

## 2 Related Work

Bug report classification helps developers understand and fix software defects. Due to the skyrocketing number of bug reports, manual classification has become time-consuming and laborious. For a long time, researchers have been exploring how to implement automatic classification of bug reports [18]. This section will summarize some existing research work.

The earliest bug classification method is Orthogonal Defect Classification (ODC) proposed by IBM's Chillarege et al. [19] in 1992. It is a method between qualitative analysis and quantitative analysis, including 13 categories (such as functions, interfaces, documents, etc.). In 2008, Antoniol et al. [6] proposed an automatic classification method for bug reports, using vector space technology to extract features, and training Decision Trees (DT), NB and LR classifiers to judge whether the report is a bug. The results show that the classification accuracy on Mozila, Eclipse and JBoss projects can reach 77% to 82%. In 2013, Pingclasai et al. [20] proposed a classification method to identify the authenticity of bugs. They adopted the topic model of Latent Dirichlet Allocation (LDA) combined with NB and Linear Logistic Regression (LLR) classifiers, and the accuracy of the three projects of HTTP-Client, Jackrabbit and Lucene reached 66% to 76%, 65% to 77% and 71% to 82%. Similarly, kukkar et al. [13] applied a hybrid method to identify whether the report is a bug or non-bug in 2019, which integrates TM, NLP and ML technologies. They observed the performance of Term Frequency-Inverse Document Frequency (TF-IDF) and feature selection and K-NN classifiers on five different data sets (Mozilla, Eclipse, JBoss, Firefox, OpenFOAM). Experiments show that the performance of the K-NN classifier varies with different data sets, and its F-measure is 78% to 96%.

In addition, there are also researchers who classify the severity of bug reports. Menzies et al. [21] presented a new automated method called SERVERS in 2008. This method uses TF-IDF, InfoGain and Rule Learning technology to divide the severity of bug reports into 5 categories, from the most severe to the most insensitive. In 2011, Sari et al. [22] applied the InfoGain method to filter out 5 valid attributes from the 14 attributes reported in the bug report for severe and non-serious classification. These 5 attributes are component, qa_contact, summary, cc_list, and product. Their combination can achieve 99.83% accuracy on the SVM model. In 2016, Zhang et al. [23] improved the REP (i.e. REP theme) and K-NN algorithm to search for historical bug reports similar to new bugs, further extracted their features to predict the severity, and classified the bug reports into Blocker, Trivial, Critical, Minor, and Major. The results show that their proposed method can effectively improve the accuracy of the prediction of the severity of bug reports. In 2019, Kukkar et al. [24] believed that traditional Machine Learning classifiers could not capture some potentially important features of bug reports, so they proposed a classification method based on Deep Learning. The model uses the N-gram algorithm and Convolutional Neural Network (CNN) and Random Forest with Boosting to solve the multi-level severity prediction problem of bug reports. Their work has achieved good results, with an average accuracy rate of 96.34% in the five open source projects.

Not only limited to bug or severity, but also researchers have proposed different classification models from other perspectives. Du et al. [25] developed an automatic classification framework based on word2vec in 2017, which classified bug reports into different fault trigger categories from four granularities, including Bug/Non-Bug, BOH/MAN, ARB/NAM, and NAM/ARB. In 2014, Tan et al. [26] believed that semantic, security and concurrency problems are strongly related to software systems. Based on the above assumptions, they studied the distribution of these three types in projects such as Apache, Mozilla and Linux, and automatically classified bug reports into the above three types through machine learning technology. The average F-measure is about 70%. Recently, Catolino et al. [27] defined a new bug report classification pattern in 2019, including 9 defect types (Configuration issue, Network issue, Database-related issue, GUI-related issue, Performance issue, Permission/deprecation issue, Security issue, Program anomaly issue, Test code-related issue). Compared with Tan et al. [26], the method of Catolino et al. can provide a clearer and comprehensive overview of the types of bug reports. At the same time, the automatic model they built also achieved higher F-Measure and AUC-ROC (64% and 74%).

It can be seen from related work that many researchers have achieved good results in the automatic classification of bug reports. On the basis of existing research, the focus of this article is to add a new factor, that is, the intention of the report, when implementing the automatic classification of bug reports. We believe that increasing this factor will improve classification performance.

## 3 Methodology

This section details the proposed classification method for bug reports. The framework is shown in Fig. 2. First, we collect and manually mark bug reports in the open repository, and then perform preprocessing steps on them. Then, we use BERT and TF-IDF methods to extract features. And the text feature and frequency feature are merged and normalized. Next, we input the extracted features into five classifiers (including K-NN, NB, LR, SVM and RF). Finally, we categorize bug reports into bug and non-bug.

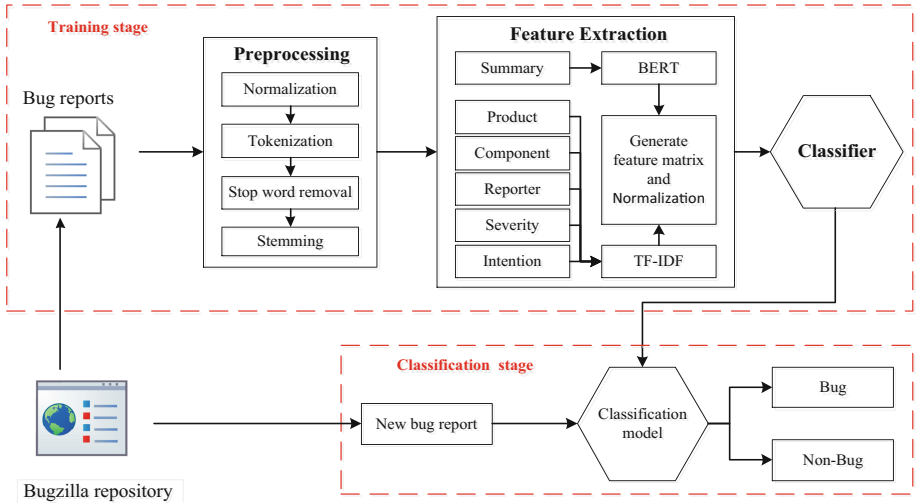

**Fig. 2.** Framework of our approach

### 3.1 Preprocessing

This experiment uses manually marked bug reports as the experimental data set. The data is input in CSV file format, and text preprocessing steps are performed on the summary field, including normalization, tokenization, stop word removal, and stemming.

(1) Normalization: Its task is to unify all words and letters in the data into lowercase.
(2) Tokenization: Its task is to delete numbers, symbols, and punctuation. In this experiment, spaces are used to replace punctuation and numbers are deleted.
(3) Stop word removal: Its task is to delete common words that do not carry specific context-related information, thereby improving the classification performance of the model.
(4) Stemming: Its task is to remove the affixes of words and extract the main part to reduce the redundancy of text data.

### 3.2 Feature Extraction

After the preprocessing step, we use BERT to extract the text features of the summary field. The BERT model is a pre-training model proposed by Google [28], which can learn dynamic context word vectors and more comprehensively capture the features of word meaning, word position and sentence meaning. In this experiment, the output of the penultimate layer of the BERT model is used as the feature score. For fields other than the summary (i.e. product, component, reporter, severity, intention), the score is calculated using the TF-IDF algorithm. The TF-IDF algorithm can indicate the importance of the field in the document, which helps to increase the classification ability of the model. Finally, the text feature scores and frequency feature scores are spliced and fused to generate a feature matrix, and normalized. The steps are shown in Table 2.

**Table 2.** The steps of feature extraction

| |
|---|
| **Input:** preprocessed bug report data |
| **Output:** feature matrix |
| **Step 1:** BERT extracts the text features of the summary field and maps the features to a high- dimensional vector matrix (M1). $$M1 = [V1, V2, \ldots \ldots Vn]$$ Vi = Feature vector, $i \in \{1,2,\ldots \ldots n\}$ |
| **Step 2:** Calculate the TF-IDF scores of other fields except the summary to obtain the matrix (M2). $$TF*IDF = Nw/N*log(D/(Dw+1))$$ Nw = The number of occurrences of the term w in the text <br> N = Total number of text entries <br> D = The total number of documents in the corpus <br> Dw = The number of documents containing the term w <br> M2 = [Tp, Tc, Tr, Ts, Ta] <br> Ti = TF-IDF score of i, $i \in \{p, c, r, s, a\}$ <br> p = Product, c = Component, r = Reporter, s = Severity, a = Intention |
| **Step 3:** The feature matrix (M) is generated by fusing text features and frequency features, and normalized. $$M = M1 + M2$$ $$M = [Tp, Tc, Tr, Ts, Ta, V1, V2, \ldots \ldots Vn]$$ Normalization method $$X' = (X - X\_min)/(X\_max - X\_min)$$ X = Original data value, X_min = Minimum value of data, X_max = Maximum value of data |
| **Step 4:** Output feature matrix. |

### 3.3 Classifier

In order to find the most suitable classifier for the proposed method, we input the extracted features into five classifiers for training respectively, and observe the performance of each classifier. These classifiers include K-NN, NB, LR, SVM and RF.

### 3.3.1 K-Nearest Neighbor

K-Nearest Neighbor is a supervised classification algorithm based on distance, which is often used in the field of data mining. The core idea is: if most of the k nearest neighbors of a sample in the feature space belongs to a certain category, the sample also belongs to this category and has the characteristics of the samples in this category. That is to say, for a given test sample and a way based on a certain distance measurement, the classification result of the current sample is predicted through the closest K training samples.

Suppose there is a training data set T = $\{(x_1, y_1), (x_2, y_2), ..., (x_N, y_N)\}$, where $x_i$ is the feature vector of the sample, and y = $\{C_1, C_2, ..., C_k\}$ is sample category, i = 1, 2, ..., N. According to the selected distance metric, find the K nearest neighbors to x in the training set T, covering the x neighborhood $N_k(x)$ of these K points. According to the index that measures the similarity between samples, the nearest K known samples of each unknown category sample are searched out to form a cluster. The voting method is used in the neighborhood to vote on the searched known samples, that is, the label category with the most occurrences among the K samples is selected to determine the category y of x:

$$y = \arg\max \sum_{x_i \in N_i(x)} I(y_i = c_j) \quad (1)$$

In Eq. (1), i = 1, 2, ..., N, j = 1, 2, ..., K. Where I is an indicator function, and when $y_i = c_j$, I is 1, otherwise it is 0.

### 3.3.2 Naive Bayes

Naive Bayes classifier is a classification technique based on Bayes' theorem. It requires that each feature used for classification is independent and does not affect each other. The core idea is to calculate the category probability of each sample, and the category with the largest probability value is used as the final classification of the sample. Suppose there is a training data set, in order to calculate the probability that the sample y classified as x. According to Bayes' theorem:

$$p(x_i|y_1, y_2, \cdots, y_n) = \frac{p(x_i)}{p(y_1, y_2, \cdots y_n)} \prod_{k=1}^{n} p(y_k|x_i) \quad (2)$$

where $p(x_i)$ and $p(y)$ represent the a priori probabilities of category $x_i$ and sample y, respectively. $p(y|x_i)$ represents the possibility that category $x_i$ is sample y, and $p(x_i|y)$ represents the possibility that sample y is category $x_i$. Usually, when we deal with classification problems, the sample contains multiple features, which can be expressed y = $(y_1, y_2, ..., y_n)$. When each feature is independent of each other, it can be known from (2):

$$p(x_i|y_1, y_2, \cdots, y_n) = \frac{p(x_i)}{p(y_1, y_2, \cdots y_n)} \prod_{k=1}^{n} p(y_k|x_i) \quad (3)$$

Regarding p($x_i$) and p($y_1$, $y_2$, ..., $y_n$) as constants, after simplifying (3), we can get:

$$x_c = \arg\max \prod_{k=1}^{n} p(y_k|x_i) \tag{4}$$

where $y_i$ is the feature of the data, and $x_c$ is the classification result of the sample. In our experiment, $y_i$ is the feature of the bug report represented by the vector, and the result of $x_c$ has two types, including bug and non-bug.

### 3.3.3 Support Vector Machine

Support Vector Machine is a classifier based on statistical learning VC dimension and structural risk minimum theory. It finds a balance between classification ability (no error classification for any sample) and model complexity (classification accuracy of a specific sample) based on limited information, with the purpose of making the classifier get the best generalization ability. Suppose there is a linear sample set ($x_i$, $y_i$), i = 1, 2, ..., n, x ∈ $R^2$, y is the category label and y ∈ {−1, 1}. The linear discriminant function in d-dimensional space is:

$$g(x) = \omega \cdot x + b \tag{5}$$

If the linear classification line can accurately separate the two types of samples, the following conditions should be met:

$$y_i = 1 \Leftrightarrow g(x_i) = \omega \cdot x_i + b \geq 1 \tag{6}$$

$$y_i = -1 \Leftrightarrow g(x_i) = \omega \cdot x_i + b \leq -1 \tag{7}$$

Simplify (6) and (7) to get:

$$y_i(\omega \cdot x_i + b) \geq 1 \tag{8}$$

At this time, the classification interval is equal to 2/‖ω‖. When the condition $y_i$(ω·$x_i$ + b) ≥ 1 is satisfied, the minimum value of ϕ(ω) = (ω·ω)/2 needs to be found. Apply Lagrange multiplier and satisfy Kuhn-Tucker conditions:

$$\alpha_i[y_i(\omega \cdot x_i + b) - 1] = 0 \tag{9}$$

Finally, the optimal classification function is obtained:

$$f(x) = \text{sgn}([\omega^* \cdot x] + b^*) = \text{sgn}\left[\sum_{i=1}^{k} a_i^* y_i(x_i \cdot x) + b^*\right] \tag{10}$$

where $\alpha_i$*, b* is the parameter to determine the optimal hyperplane, and ($x_i$·x) is the inner product of the two vectors.

### 3.3.4 Logistic Regression

Logistic Regression, also known as logistic regression analysis, is a generalized linear regression classification model, which is often used in the field of data mining. Logistic regression is essentially a binary classification problem, and its dependent variable Y has two values {0, 1}. The formula of the multiple logistic regression classifier is as follows:

$$\pi(X_1, X_2, \ldots, X_n) = \frac{e^{Y_0 + Y_1 \cdot X_1 + \cdots + Y_n \cdot X_n}}{1 + e^{Y_0 + Y_1 \cdot X_1 + \cdots + Y_n \cdot X_n}} \tag{11}$$

where Xi is a vector describing the features of the data, and $1 \geq \pi \geq 0$ is the value on the logistic regression curve. In order to achieve classification, it is also necessary to set a threshold. For example, the threshold value in the model is 0.5, and x represents the text feature and frequency feature extracted from the bug report. When $\pi > 0.5$, the report is classified as bug; when $\pi < 0.5$,, the report is classified as non-bug.

### 3.3.5 Random Forest

Random Forest is a classification algorithm based on ensemble learning method, and its basic unit is decision tree. It includes "random" and "forest" parts. "Forest" means that the classifier consists of many trees, and it is based on ensemble learning theory. Random includes two aspects: one is for the training process. In order to ensure that all samples have a chance to be drawn once, the classifier randomly selects a training sample set, and the data used in each round of training is randomly selected from the original sample set with replacement. The other is for feature selection. Assuming that the original data has M features, S number of features are randomly selected from M features as candidate features of the training tree. After the training samples and features are determined, a decision tree is constructed on each training sample to get the prediction result. N samples can get N prediction models, and then use the model to predict the test samples, so that each sample can get N prediction results, and finally determine the final result through a simple majority voting principle. The formula of the model is as follows:

$$H(S) = \text{argMax} \sum_{i}^{n} I(h_i(S_i) = Y) \tag{12}$$

where $h_i(S_i)$ is a single decision tree, Y is the prediction result, and I is an indicator function.

## 4  Experiment

This experiment divides the training data and test data by 8:2, and extracts the features of the report summary field, other fields (product, component, reporter, severity), and intention, respectively. In order to find the most suitable classifier for the proposed method, we superimpose and fuse these three features in turn, and input them into five different machine learning classifiers (K-NN, NB, SVM, LR, RF) for experiments.

The experiments solved the following research questions:

RQ.1 Does adding the intention of the report improve the accuracy of the automatic classification for bug reports?

RQ.2 How about the performance of our proposed method on five different classifiers?

### 4.1 Dataset

In this study, we collected 2,230 bug reports from four ecosystems in the Bugzilla repository, respectively from Apache [14], Eclipse [15], Gentoo [16] and Mozilla [17]. Specifically, we select the reports whose status is "RESOLVED" or "VERIFIED" and the resolution is "FIXED". And extract their product, component, reporter, severity, and summary tags. On this basis, we manually marked the types and intention of these reports, and their type information statistics are shown in Table 3.

**Table 3.** Type statistics of our dataset

| Ecosystem | Total | Bug | Non-Bug |
|---|---|---|---|
| Apache | 446 | 296 | 150 |
| Eclipse | 658 | 419 | 239 |
| Gentoo | 511 | 294 | 217 |
| Mozilla | 615 | 425 | 190 |

### 4.2 Evaluation Metrics

To measure the performance of classification, we use Accuracy, Precision, Recall and F-Measure. Their definitions are as follows:

$$Accuracy = \frac{TP + TN}{TP + TN + FP + FN} \quad (13)$$

$$Precision = \frac{TP}{TP + FP} \quad (14)$$

$$Recall = \frac{TP}{TP + TN} \quad (15)$$

$$F - measure = 2 \times \frac{Precision \times Recall}{Precision + Recall} \quad (16)$$

where TP is the number of true positives, TN is the number of true negatives, FP is the number of false positives, and FN is the number of false negatives. In order to deal with the randomness caused by different data splits, ten-fold cross-validation is used to obtain the average value of evaluation metrics to measure the performance of classification.

### 4.3 Results

**RQ1. Does Adding the Intention of the Report Improve the Accuracy of the Automatic Classification for Bug Reports?**

We use three types of features that are sequentially fused and superimposed to train the classifier, and the average accuracy of the ten-fold cross-validation is shown in Table 4. Text represents the textual feature of the summary, Freq represents the word frequency feature of other fields (product, component, reporter, severity), and Intention represents the feature of the intention of the bug report. The unit of the values in the table is the percentage system. Text+Freq+Intention is the method we put forward.

Table 4. Average accuracy of all datasets

| Ecosystem | Features | Classifier | | | | |
|---|---|---|---|---|---|---|
| | | K-NN | NB | LR | SVM | RF |
| Apache | Text | 60.5 | 65.5 | 65.6 | 66.4 | 63.0 |
| | Text+Freq | 70.6 | 80.0 | 70.9 | 70.9 | 85.7 |
| | Text+Freq+Intention | **90.4** | **89.2** | **90.8** | **91.0** | **91.7** |
| Eclipse | Text | 61.5 | 63.7 | 65.0 | 64.6 | 61.0 |
| | Text+Freq | 66.4 | 66.1 | 65.2 | 64.4 | 73.1 |
| | Text+Freq+Intention | **83.9** | **84.0** | **84.8** | **84.8** | **84.8** |
| Gentoo | Text | 67.7 | 61.8 | 57.3 | 62.8 | 67.3 |
| | Text+Freq | 83.2 | 73.6 | 71.2 | 72.8 | 87.3 |
| | Text+Freq+Intention | **91.8** | **85.1** | **86.1** | **87.7** | **94.5** |
| Mozilla | Text | 65.2 | 66.8 | 65.0 | 69.4 | 67.5 |
| | Text+Freq | 75.3 | 70.4 | 72.0 | 72.3 | 78.2 |
| | Text+Freq+Intention | **89.9** | **87.5** | **87.8** | **88.0** | **87.8** |

**RQ2. How About the Performance of Our Proposed Method on Five Different Classifiers?**

The performance of our proposed method, which combines text, frequency and intention features (Text+Freq+Intention), on the five classifiers is shown in Figs. 3, 4, 5, 6 and 7. The x-axis represents the source of the data, and the y-axis represents the average value of the ten-fold cross-validation.

142    F. Meng et al.

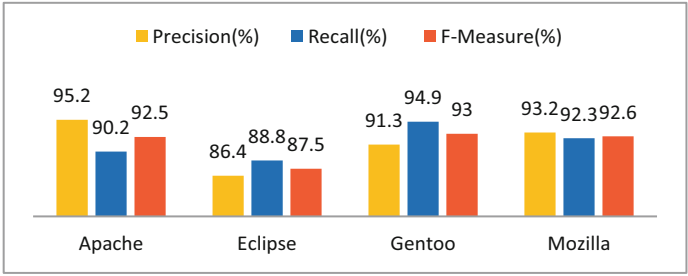

**Fig. 3.** The performance of all data on the K-NN classifier

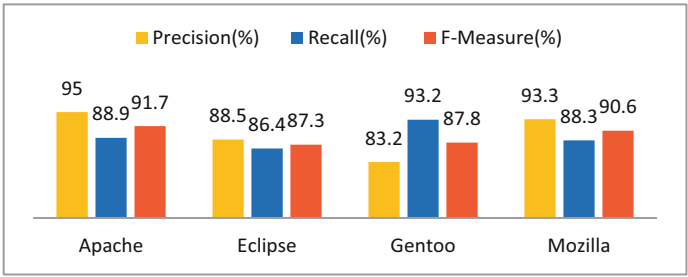

**Fig. 4.** The performance of all data on the NB classifier

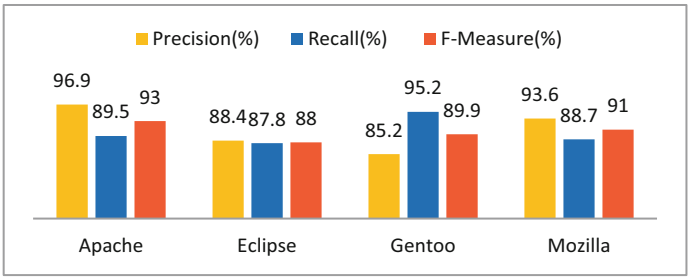

**Fig. 5.** The performance of all data on the SVM classifier

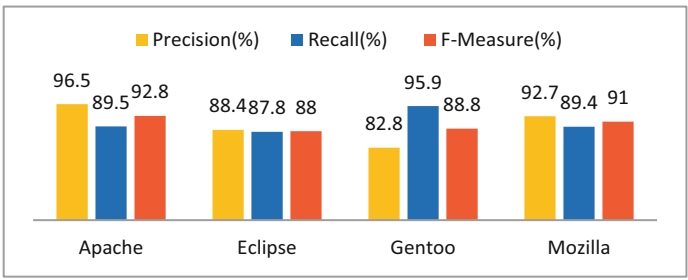

**Fig. 6.** The performance of all data on the LR classifier

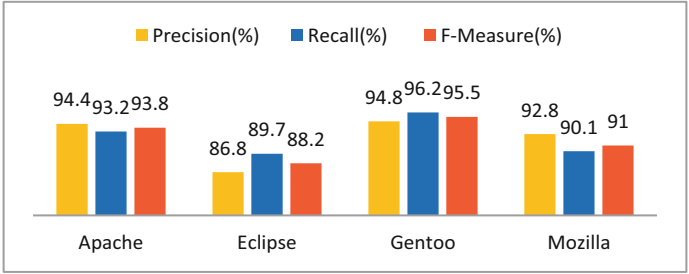

**Fig. 7.** The performance of all data on the RF classifier

## 5  Discussion

### 5.1  Experiment Analysis

Performance of our method: we combine the proposed approach with five different machine learning classifiers, and conduct experiments on the datasets of Apache, Eclipse, Gentoo, and Mozilla. Table 4 shows the average Accuracy of all data sets on different classifiers. Among them, the Apache data set has a maximum of 91.7%, the Eclipse data set has a maximum of 84.8%, the Gentoo data set has a maximum of 94.5%, and the Mozilla data set has a maximum of 89.9%. As can be seen from Table 4, compared with considering the text field of the report alone, after adding the intention factor of the report we proposed, the accuracy of the data sets of the four ecosystems on the five classifiers has been significantly improved. To explore why adding the binary feature of reporting intention can improve classification performance, we made statistics on the distribution of intention features and their correlation with labels (i.e. bug or non-bug) in the experimental dataset. The results are shown in Table 5.

**Table 5.** Distribution statistics of intention features for all data

| Ecosystem | Type | Intention distribution | |
|---|---|---|---|
| | | Explanation | Suggestion |
| Apache | Bug | 265 | 31 |
| | Non-bug | 9 | 141 |
| Eclipse | Bug | 368 | 51 |
| | Non-bug | 49 | 190 |
| Gentoo | Bug | 284 | 10 |
| | Non-bug | 66 | 151 |
| Mozilla | Bug | 373 | 52 |
| | Non-bug | 23 | 167 |

From the above table, it can be concluded that there is an imbalance in the distribution of intention features in the dataset. Across the four ecosystems, bug reports included more explanations than non-bug reports, and non-bug reports included more suggestions. The imbalanced distribution of binary features can make the classifier more sensitive to different labels during training. Therefore, adding a binary feature of reported intention can improve classification performance.

In addition, in order to test the scalability of our method and select a classifier suitable for it, we also tested the performance of our approach on K-NN, NB, LF, SVM and RF classifiers. Figures 3, 4, 5, 6 and 7 shows the evaluation index value of each classifier. The results show that our proposed method has achieved good results in Precision, Recall and F-measure on five classifiers. In all data sets, the Precision reached 82.8% to 96.9%, the Recall reached 86.4% to 96.2%, and the F-measure reached 87.3% to 95.5%. Among the five classifiers, the comprehensive performance of Random Forest is better than other classifiers, and the performance of each classifier will change with the different data sets. Although the performance of our method changes with the classifier, the overall result is still a good level. This also means that when using our method to conduct a bug report classification, you can adjust the classifier according to different ecosystems and task requirements to achieve the best results. We believe our method is scalable.

Some of our thoughts:

(1) Bugs are harder to understand than non-bugs. Therefore, when faced with bug-type defects, only a few reporters can provide solutions, and most reporters can only describe the problem. This results in most reports where the intention is "explanations" are bugs, and those where the intention is "suggestions" are non-bugs.
(2) In the Bug Tracking System of open source software, the reporters are not only software developers and testers, but also a large number of users. The reporters who explain the defects are mostly users, and the reporters who can make suggestions for the defects are mostly software developers and testers. Because developers and testers have richer experience and knowledge than users, they have a better understanding of the code and architecture of the program, and can give advice on complex defects.
(3) The intention labels in this study are manually labeled, and this work seems to increase the training time of the automatic classification model, but we are to verify that the proposed method can improve the performance of automatic classification of bug reports. We think it is possible to add the label of reporting intention to the Bug Tracking System, so that reporters can explain their intentions when submitting reports, which can greatly reduce the time for manual labeling during model training. It is much easier for reporters to state their intentions than to judge whether a report is a bug.

### 5.2 Threats to Validity

In this part, we identified the following threats that may exist in this study.

Internal threat: The bug report tags of most open repositories contain errors. In order to avoid incorrect labeling to affect the performance of the model, the data set of this experiment is constructed by our manual labeling based on the data in the Bugzilla repository. Although we have flagged bugs or non-bugs in accordance with the rules proposed in the existing literature [29], due to differences in experience and knowledge background, there may be flagging errors that affect the performance of the model.

External threat: This research focuses on the 2,230 bug reports of the four ecosystems (Apache, Eclipse, Gentoo, Mozilla) to classify bugs or non-bugs. However, the performance of our method on other ecosystems is unknown, that is, the performance of our bug report classification model on data from other software systems may be higher or lower than the results of our experiments.

## 6  Conclusion and Future Work

In this study, we propose a new automatic classification approach for bug reports, that is, to increase the intention of the report based on the text information of the report. Our approach combines Text Mining, Natural Language Processing and Machine Learning technologies. We first collected 2,230 reports from the four ecosystems (Apache, Eclipse, Gentoo, Mozilla) in the bug repository, and manually marked their types and intention, with the goal of constructing the data set required for the research. Then, we perform preprocessing steps on the data, extract the text features of the report summary field and the word frequency features of other fields, and add the intention features of the report. Next, we superimpose these features and input them into five classifiers (K-NN, NB, SVM, LF, RF). Finally, we classify bug reports into bugs and non-bugs. The results show that, compared with simply extracting text information features for classification, adding the intention features of the report we proposed can significantly improve the performance of bug report classification. In the future, we will verify the proposed approach on more open source ecosystems, and combine Deep Learning technology to improve the performance of automatic classification of bug reports.

**Acknowledgment.** This work is supported by the Science and Technology Research Project of the Jilin Provincial Department of Education, "Research on Overtime Risk Assessment and Early Warning Technology of Industrial Control Code" (No. JJKH20210097KJ).